\documentclass{article}

\usepackage[a4paper,margin=1.7in]{geometry}

\usepackage{lipsum}
\usepackage{amsfonts}
\usepackage{amsfonts}
\usepackage{amsthm}
\usepackage{amsmath}
\usepackage{graphicx}
\usepackage{epstopdf}
\usepackage{algorithmicx}
\usepackage{algorithm}
\usepackage{algpseudocode}
\DeclareGraphicsExtensions{.eps,.pdf,.png,.jpg}
\usepackage{amsopn}

\title{Tomographic Reconstruction Methods for Decomposing Directional Components}
\author{Rasmus Dalgas Kongskov\footnote{Department of Applied Mathematics and Computer Science, Technical University of Denmark, 2800 Kgs. Lyngby, Denmark (rara@dtu.dk, yido@dtu.dk.} \and Yiqiu Dong\footnotemark[1]}

\usepackage{color}
\definecolor{pRed}{rgb}{0.7,0.01,0.15}
\definecolor{pBlue}{rgb}{0.01,0.01,0.8}
\definecolor{pGreen}{rgb}{0.01,0.8,0.01}
\definecolor{dGreen}{rgb}{0.01,0.4,0.01}

\usepackage{hyperref}
\hypersetup{linkcolor=dGreen,citecolor=dGreen,urlcolor=dGreen}

\graphicspath{{figs/}}

\def\NoNumber#1{{\def\alglinenumber##1{}\State #1}\addtocounter{ALG@line}{-1}}

\newcommand{\figone}[3]{
\begin{figure}[h] 
\centerline{\includegraphics[width=#2\columnwidth]{#1}}
\caption{#3} \label{fig:#1}
\end{figure} }

\usepackage{cleveref}
\crefformat{equation}{(#2#1#3)}

\newcommand{\TV}{\mathrm{TV}}			
\newcommand{\DTV}{\mathrm{DTV}}		

\newcommand{\objf}{z} 				
\newcommand{\primf}{u} 				
\newcommand{\secf}{w} 				
\newcommand{\objfd}{\mathbf{\objf}} 	
\newcommand{\primfd}{\mathbf{\primf}} 	
\newcommand{\secfd}{\mathbf{\secf}} 	
\newcommand{\sysm}{A}					
\newcommand{\noisef}{b}				
\newcommand{\sino}{\mathbf{\noisef}}	
\newcommand{\sparse}{\beta}			
\newcommand{\widE}{a}					
\newcommand{\angE}{\theta}			
\newcommand{\sysmA}{\sysm_\primf}		
\newcommand{\sysmB}{\sysm_\secf}		
\newcommand{\sinoA}{\sino_\primf}		
\newcommand{\sinoB}{\sino_\secf}		
\newcommand{\main}{h}					
\newcommand{\ctang}{\phi}			    
\newcommand{\ctangA}{\ctang_\primf}	
\newcommand{\ctangB}{\ctang_\secf}	
\newcommand{\freqsum}{W}				
\newcommand{\fft}{\hat{\noisef}}		
\newcommand{\scoord}{x} 				
\newcommand{\regup}{\lambda}			
\newcommand{\reguR}{\mathcal{R}}		
\newcommand{\reguw}{\alpha}			
\newcommand{\rot}{R}					
\newcommand{\scm}{\Lambda}			
\newcommand{\xind}{i}					
\newcommand{\yind}{j}					
\newcommand{\tind}{l}					
\newcommand{\thind}{m}				
\newcommand{\forw}{+}					
\newcommand{\pdim}{M}					
\newcommand{\sdim}{N}					
\newcommand{\ctbin}{t}				
\newcommand{\splitR}{K}				
\newcommand{\noisel}{\rho}			
\newcommand{\radon}{\mathcal{A}}		
\newcommand{\fdm}{\nabla}				
\newcommand{\exvec}{\mathbf{y}}		
\newcommand{\elvec}{y}				

\theoremstyle{plain}

\theoremstyle{definition}

\begin{document}

\maketitle

\begin{abstract}
Decomposition of tomographic reconstructions has many different practical application. We propose two new reconstruction methods that combines the task of tomographic reconstruction with object decomposition. We demonstrate these reconstruction methods in the context of decomposing directional objects into various directional components. Furthermore we propose a method for estimating the main direction in a directional object, directly from the measured computed tomography data. We demonstrate all the proposed methods on simulated and real samples to show their practical applicability. The numerical tests show that decomposition and reconstruction can combined to achieve a highly useful fibre-crack decomposition.
\end{abstract}

\section{Introduction}

X-ray Computed Tomography (CT) is a highly used non-invasive imaging technique. Applications of this technique ranges from biological and chemical science, to structural and material science, where the resolution also varies from large scale (meters) to micro-scale (nano-meters). The technique is based on the attenuation that X-rays undergoes as they pass trough matter. The attenuated X-ray intensity is measured at a detector. In this paper our CT set-up is 2D and parallel beam, but the methods introduced here are not limited to this CT-methodology, but presented in this setting for simplicity. The X-ray source and detector is rotated around the object, and for each angle of rotation the attenuated X-ray signal is measured.

Lambert-Beers law relates the object that we are scanning to the measured intensity. By rewriting Lambert-Beers law, the measured intensity is seen to be linearly related to the object through line-integrals \cite{buzug2008}. If we model the object as a continuous function the relation between the measured intensity and object corresponds to the Radon transform. In this paper we focus on the discretized version of this model where the domain with the object, the reconstruction domain, is split into $\pdim\times\pdim$ equidistant pixels. The discrete model can be expressed as a simple matrix-vector product as 
\begin{align} \label{eqn:discCT}
\sysm \objfd = \sino.
\end{align}
Here the system matrix, $\sysm\in\mathbb{R}^{\sdim\times\pdim^2}$, models the length of each X-ray through each pixel in the reconstruction domain. The reconstruction, $\objfd\in\mathbb{R}^{\pdim^2}$, is the object, including it's interior, that we seek to find. The sinogram data, $\sino\in\mathbb{R}^N$, is the measured data at the detector. The data is measured at a series of discrete angles, $\ctang\in\mathbb{R}^{\sdim_\ctang}$, where $\sdim_\ctang$ is the number of discrete angles. For simplicity we assume that we have parallel beam measurements on a 180$^\circ$ degree arch, i.e. not \textit{limited angle} measurements. We further assume that we do not have \textit{sparse angle} measurements. Here sparse angle measurements refers to the case where the amount of measurement angles is significantly smaller than the amount of detector bins $\sdim_\ctbin$. In order make the dimensions fit we have $\sdim=\sdim_\ctang\sdim_\ctbin$. The application of $\sysm$ is usually called a forward projection, whereas the application of $\sysm^T$ is a called a backward projection.

The process of constructing $\objfd$ based on known $\sino$ is usually called a \textit{reconstruction}. The most widely used reconstruction method for CT-problems was introduced in \cite{Radon17} and is now known as Filtered Back-Projection (FBP). The FBP method implicitly assumes to have continuously measured data from the whole 180$^\circ$ angular range. For a discrete problem this assumption is violated, but with sufficiently many equidistant angular measurements, and a low no noise level, this method will give results close to the exact object. The main advantage of FBP is the efficiency since the computational effort of this algorithm relies on a two Fourier transforms, which numerically can be handled by use of the fast Fourier transform (fft). The disadvantage of FBP is that the method is sensitive toward measurement noise and any type of limited data problems.

Variational methods have been used in many types of inverse problems, see \cite{Aubert2006,Scherzer2010} for more on this topic. The main reason for using variational methods, is that can help us to overcome unwanted the effects of ill-posedness, noise and limited data. Another advantage of using variational methods is the possibility for including prior knowledge through regularization.  In this paper we will make use of variational methods for solving the reconstruction problem, which will be on the form
\begin{align} \label{eqn:varprop}
\min_\objfd \ \frac{1}{2}\| \sysm \objfd - \sino \|^2_2 + \reguR(\objfd).
\end{align}
The first term, the \textit{data-fidelity term}, is chosen to be the 2-norm since we model noise as being Gaussian. The second term is the \textit{regularization term} and here we will use several different methods. Typically a \textit{regularization parameter}, that balances the two terms, is also present in a variational formulation, here we include this parameter as part of the regularization term. One of the most widely used regularization methods for imaging problems is total variation (TV), first introduced in \cite{Rudin1992} and used for tomography problems in e.g. \cite{Delaney1998,Jonsson,Sidky2008}.

After reconstructing an object one is typically interested in analyzing the result, which could be aided by the use of segmentation. If we expect that our object consist of several distinct features that are mixed together, one could consider to decompose the image, i.e. split the object into different parts. In order to decompose an object a decomposition model that describes how the object should be split is needed. In \cite{Chambolle1997} a way to decompose a signal using \textit{infimal convolution} is introduced. For two convex functionals $J_1$ and $J_2$ a function is linearly split into two components as $\objf = \objf_1 + \objf_2$ and the infimal convolution of the two functionals becomes
\begin{align*}
J(\objf) = \inf \left\{ J_1(\objf_1) + J_2(\objf_2) \right\}.
\end{align*}
This strategy relies on the assumption that each component, $\objf_1$ and $\objf_2$, attains a smaller functional value for the functional which models the given component. Meyer further investigated image decomposition in \cite{Meyer2001}, where methods for modelling and extraction of oscillating patterns in images was investigated in great depth. Theese earlier works sparked research of image decomposition methods in various aspects, see e.g. \cite{Esedoglu2004,Starck2005,Aujol2005}. Variational methods for inverse problems is very closely related to the decomposition methods and in \cite{Aujol2006,Gilles2007} denoising was combined with image decomposition to texture and structure parts. The idea of using image decomposition for inverse problems is to split an object linearly into two components, $\objf = \primf + \secf$, and let the regularization consist of two different terms, one for each component. In the more recent work \cite{Holler2014} infimal convolution was shown to give good results when combining several Total Variation (TV) type functionals. 

Solving \cref{eqn:varprop} for $\objfd = \primfd + \secfd$ and different regularization terms for $\primfd$ and $\secfd$, will lead to a solution where the two components will follow different priors, and hence a reconstruction and a decomposition is achieved simultaneously. Other attempts of decomposing CT reconstructions has been through segmentation, either as post-processing, see e.g. \cite{Rouse2012}, or simultaneous with the reconstruction, see e.g. \cite{Romanov2016}. Decomposition of an object by use of spectral CT has also been attempted, see \cite{Long2014}. In \cite{Li2016} a method for simultaneously decomposing and reconstructing an object is introduced. In this paper the object is decomposed into a three components: the ground truth object, the limited data artifacts and the measurement noise, which are assumed to be sparse in the wavelet basis, the discrete cosine transform basis and in the sinogram domain, respectively.

Directional objects are objects where the texture in the object follows one main direction. Applications with directional objects are, among other things, fibres, such as optical fibres, glass fibres, carbon fibres, etc. When analyzing fibre materials, CT scanners can be used to investigate interior properties, for example irregularities, see \cite{Rouse2012,Sandoghchi2014,Jespersen2016}. A specific irregularity which is often sought for in fibre materials are cracks which are distinctive from the fibres. Both the fibres and the cracks can be regarded as directional components, that are part of an object. Recently a regularization method for directional objects has shown promising results in \cite{Kongskov2017}. The first order regularization method presented in this paper is based on TV and hence it is called Directional Total Variation (DTV). 

In this paper we propose two combined reconstruction and decomposition methods that benifits from using a variational formulation. The one reconstruction method is based on the infimal convolution decomposition method, where solving the variational minimization problem decomposes the object into two, or more, components. The other reconstruction method is motivated from the microlocal analysis results in \cite{Quinto1993}, and utilizes regularization through a variational formulation to overcome consequential limited data artifacts. We introduce both reconstruction methods in the light of reconstructing directional objects, and we demonstrate advantages and disadvantages of the reconstruction methods through empirical examples. In order to use DTV regularization for the CT reconstruction problem we need to supply the main object direction. We therefore propose a method for estimating the main objection direction directly from sinogram data. A practical application of the directional decomposition methods could be a fibre material with cracks that occur across the main object direction, i.e. perpendicular to the direction of the fibres.

The paper is organized as follows. 
In \cref{sec:DTVCT} directional regularization for CT problem is introduced and demonstrated. Furthermore we give a method for estimating the main direction of an object directly from the measured sinogram data. 
In \cref{sec:sinogramsplit} we introduce the \textit{sinogram splitting} method, where the entire reconstruction problem is split into two individual problems. 
In \cref{sec:dtvdecom} \textit{DTV-decomposition}, a variational decomposition model based on two different DTV regularization term, is introduced. 
Both sinogram splitting and DTV-decomposition makes use of the direction estimation method and DTV, but one method is obviously advantageous to the other and we therefore introduce both method.
Numerical experiments are carried out in \cref{sec:numexp}, where the two introduced decomposition methods are tested and compared. 
In \cref{sec:conclusion} conclusions are drawn.

\section{Directional regularization for CT problems} \label{sec:DTVCT}

We introduce a regularization method for CT problems where the object of interest is directional, i.e. the texture of the object follows one specific main direction. Furthermore we propose a method for estimating the main direction of an object directly from the sinogram data.

The domain consist of $\pdim$-by-$\pdim$ pixels with equidistant pixel-grid-size $1\times1$. In this region $(\xind,\yind)$ denotes a pixel index with $1\leq\xind, \yind\leq M$ such that $\primf_{\xind,\yind}$ gives the pixel value at $(\xind,\yind)$.
Since the scanning is circular, the region of interest, and our simulated phantoms, are contained in a circle with the diameter $\pdim$.
The sinogram domain is discretized in $\sdim_\ctbin$-by-$\sdim_\ctang$, so we have $\sino\in \mathbb{R}^{\sdim_\ctbin\times\sdim_\ctang}$. In the sinogram domain $(\tind,\thind)$ denotes the pixel index such that $\noisef_{\tind,\thind}$ is sinogram value at detector bin $\ctbin_\tind$ and measurement angle $\ctang_\thind$.

\subsection{Directional regularization}
Regularization methods has been used for CT problems for almost 40 years, see \cite{Levitan1979,Louis1983} for some of the first works in this direction. Over time numerous regularization techniques has been adapted, developed and utilized for the object-reconstruction process. Since the publication of the paper that introduced the ROF-model for imaging problems \cite{Rudin1992}, Total Variation (TV) has been used for a wide range of imaging applications. In discrete terms the total variation of $\objfd\in\mathbb{R}^{\pdim\times\pdim}$ can be expressed as
\begin{align*}
\TV(\objfd)  = \sum_{i,j} |(\fdm \objfd)_{i,j} |_2,
\end{align*}
for euclidian norm $|\cdot|_2$ and the discrete gradient operator $\fdm:\mathbb{R}^{\pdim\times\pdim}\to\mathbb{R}^{2\pdim\times\pdim}$. $\fdm$ is defined as
\begin{align*}
\fdm\objfd=\begin{pmatrix} \fdm_{\scoord_1}\primf \\ \fdm_{\scoord_2}\primf \end{pmatrix},
\end{align*}
where $\nabla_{\scoord_1}$ and $\nabla_{\scoord_2}$, for the two dimensions $\scoord_1$ and $\scoord_2$, are obtained by applying a forward finite difference scheme with symmetric boundary condition, i.e., \begin{align*}
(\fdm^\forw_{\scoord_1}\objfd)_{\xind,\yind} = \left\{ \begin{array}{ll} \primf_{\xind+1,\yind}-\primf_{\xind,\yind}, & \text{if}\; \xind<\pdim, \\ 0, & \text{if}\; \xind=\pdim, \end{array} \right. \, \text{and} \,
(\fdm^\forw_{\scoord_2}\objfd)_{\xind,\yind} = \left\{ \begin{array}{ll} \primf_{\xind,\yind+1}-\primf_{\xind,\yind}, & \text{if}\; \yind<\pdim, \\ 0, & \text{if}\; \yind=\pdim. \end{array} \right.
\end{align*}

A regularization method for processing images while incoporating directional priors has been introduced for image denoising and image deblurring problems in \cite{Kongskov2017,Kongskov2017b}. The first order regularization method, DTV, builds on the prior that the target object is piece-wise constant with the texture along one main direction. In this paper we will use DTV for regularizing the CT-reconstruction problem. In discrete terms we can express DTV as 
\begin{align*}
\DTV_{\angE,\widE}(\objfd)  = \sum_{i,j} |\rot_\angE\scm_\widE(\fdm \objfd)_{i,j} |_2,
\end{align*}
for scaling and rotation matrices
\begin{align} \label{eqn:dtvmat}
\rot_\angE = \begin{pmatrix} \cos\angE & -\sin\angE \\ \cos\angE & \sin\angE \end{pmatrix} \quad \mathrm{and} \quad \scm_\widE = \begin{pmatrix} 1 & 0 \\ 0\ & \widE \end{pmatrix},
\end{align} 
where $\widE\in (0,1]$ and $\theta\in (0,2\pi]$.

Fibre materials are objects that are often examined using CT, and these fit well with the prior for DTV, since they are piecewise constant have highly uni-directional texture. We therefore center our examples around fibre materials or simulated fibre materials. We give an example where the aim is to reconstruct a simulated directional phantom. In this example we are dealing with a slightly underdetermined system, $\frac{2}{3}$, with 1\% Gaussian sinogram-noise. We compare the results produced by FBP, $\ell_2$-$\TV$ and $\ell_2$-$\DTV$ by visualizing the results and by means of the peak signal-to-noise ratio (psnr) measure in \cref{fig:comp3}. By visual and quantitative comparison we see the advantage of incorporating the additional prior for fibre objects.

\figone{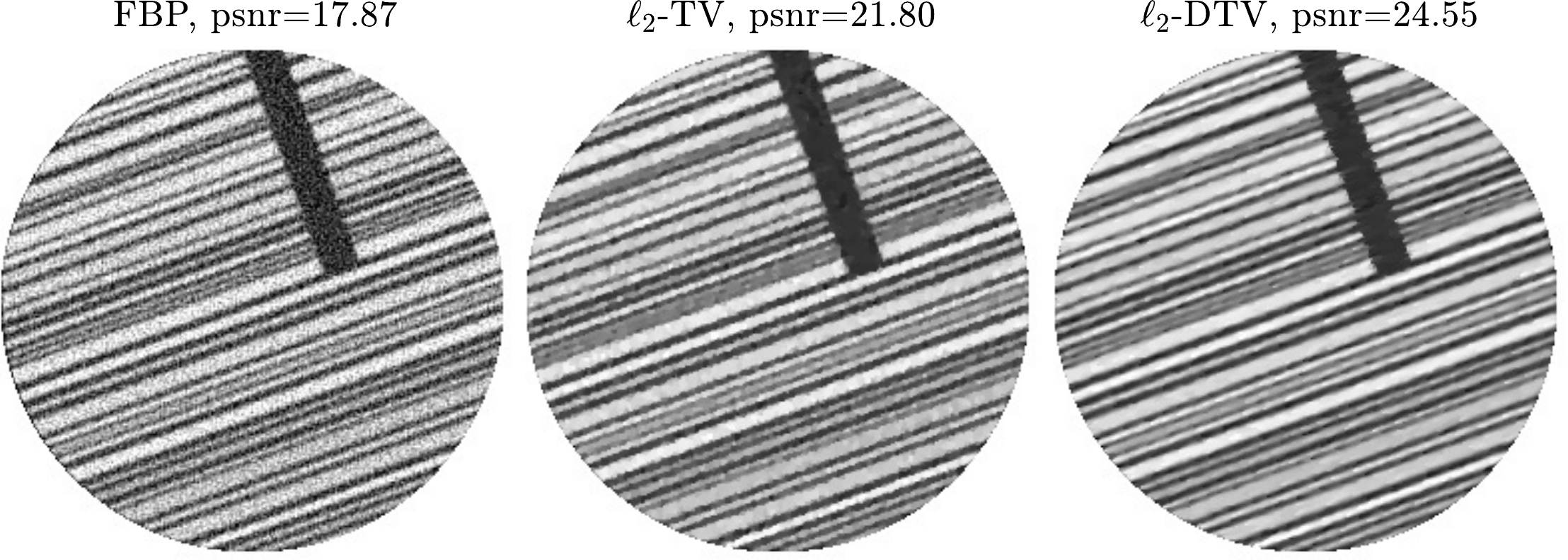}{0.8}{Simulated CT-problem solved using three different reconstruction techniques in order to compare their performance. Regularization parameters for the $\ell_2$-TV and $\ell_2$-DTV methods are tuned to maximize peak-signal-to-noise (psnr) ratio. DTV parameters are chosen as $\widE=0.15$ and $\angE=\frac{\pi}{9}$ $(20^\circ)$.}

In order to regularize an inverse problem using DTV, the main direction should be know. Next we introduce a method for estimating the main direction directly from the measured CT-data. Furthermore a width-parameter $\widE$ should be chosen, but based on the conclusions from empirical tests in \cite{Kongskov2017} we just set $\widE=0.15$ in this paper.


\subsection{Direction estimation from CT-data} \label{sec:EstDir}

In order to use DTV for regularization we have to select the parameters $\widE$ and $\angE$. 
From \cite{Kongskov2017} we know that if the directional prior is met we should choose $\widE\in ]0,1]$ as small. 
Empirical tests in \cite{Kongskov2017} show that $\widE=0.15$ is a reasonable choice for images, we choose this value for the CT reconstruction problem.

In \cite{Kongskov2017b} a method for finding the main direction in corrupted images, using the Fourier transform, was presented. This method relies on Fourier transforming the corrupted image and analyzing the magnitude of the coefficients. If the image texture has a main direction, the coefficients corresponding to basis-functions with same main direction, will be relatively large in magnitude. 

To find the main-direction $\angE$ in the object of interest we can utilize that CT data already consist of angular measurements. Given that we have data measured from one or more angles relatively close to the main direction of the object, we can detect this main angle by comparing the individual angular measurements. Inspired by the method in \cite{Kongskov2017b} our method consist of a Fourier transform, though in 1D, for each angle, i.e. along the detector-pixels. We sum the magnitude of the Fourier coefficients, for each angle, and find the maximum response. The algorithm is outlined in \cref{alg:estang}. Data measured at the same angle as the main direction will be oscillatory. An oscillating signal is well represented by the Fourier basis-functions, and therefore we expect the magnitude of the Fourier coefficients to be largest at one specific angle, which will therefore become the estimated main direction. 

\begin{algorithm}
\caption{Main Direction Estimation Method}
\label{alg:estang}
\begin{algorithmic}[1]
\vspace{1mm}
\State Input siogram data $\noisef$ and discrete measurement angles $\ctang$. \vspace{1mm}
\NoNumber{1D Fourier transform along each angle $\ctang_\thind$:} \vspace{1mm}
\NoNumber{$\fft_{\omega,\thind} = \sum_{\tind=0}^{\sdim_\ctbin-1} \  \noisef_{\tind,\thind} \mathrm{e}^{-\frac{2\pi\iota\omega\tind}{\sdim_\ctbin}} $. \vspace{1mm} }
\State Calculate summed magnitude of $\fft$ for each angle $\ctang_\thind$ and find max response: \vspace{1mm}
\NoNumber{$ \freqsum_\thind = \underset{\omega}{\sum}\ |\fft_{\omega,\thind}| $,}
\NoNumber{$\main = \underset{\thind}{\text{argmax}}\ \freqsum_\thind $.}
\State $\angE = \ctang_\main$. \vspace{1mm}
\NoNumber{\Return $\angE$.}
\end{algorithmic}
\end{algorithm}

It should be noted that this angle estimation method is not limited to parallel-beam tomography since a pattern also occurs when the fan-beam geometry is used, although the data should be re-binned or the method should be modified slightly to account for this.

\begin{figure} 
\centering
\includegraphics[width=0.36\textwidth]{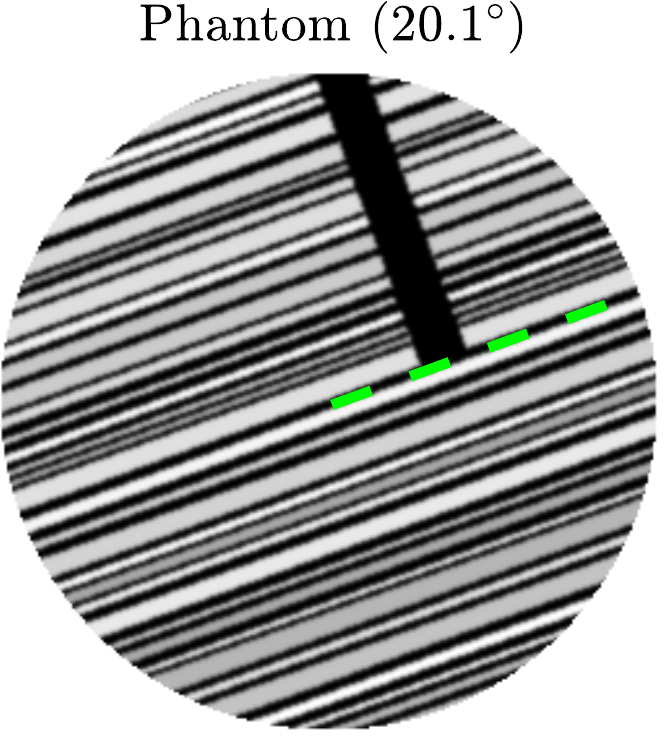}
\hspace{0.05\textwidth}
\includegraphics[width=0.53\textwidth]{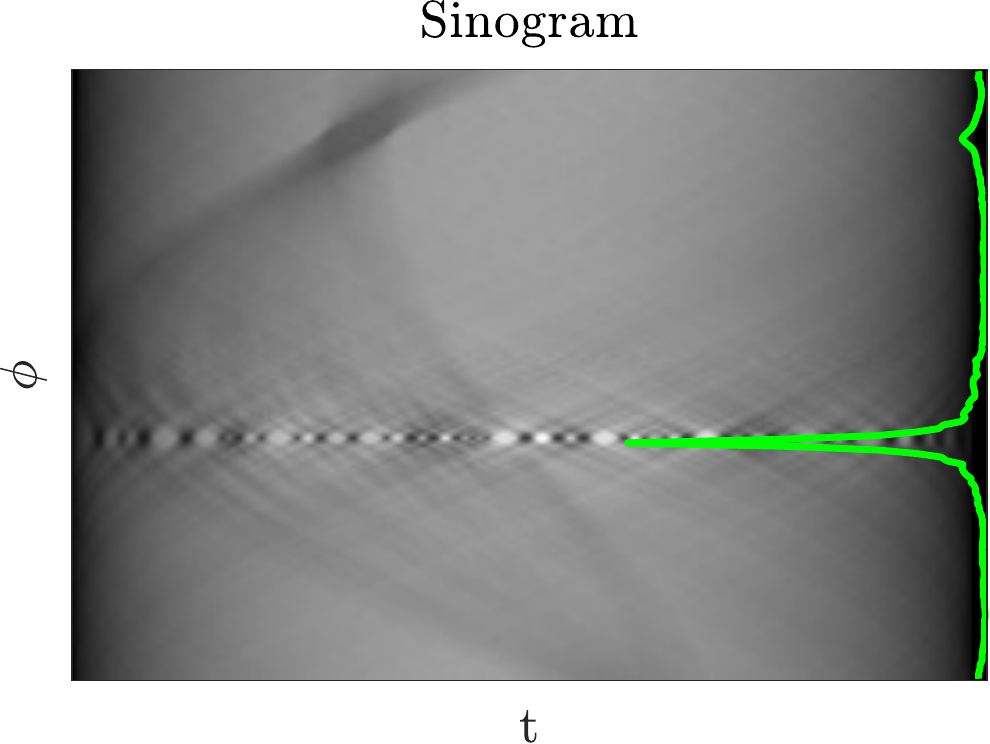}
\caption{To the left: Fibre-phantom with estimated main direction marked by dashed line. To the right: Noise-free sinogram overlayed with a line-plot of $\freqsum$ (see \cref{alg:estang}).} \label{fig:dir_est}
\end{figure}

\begin{figure} 
\centering
\includegraphics[width=0.36\textwidth]{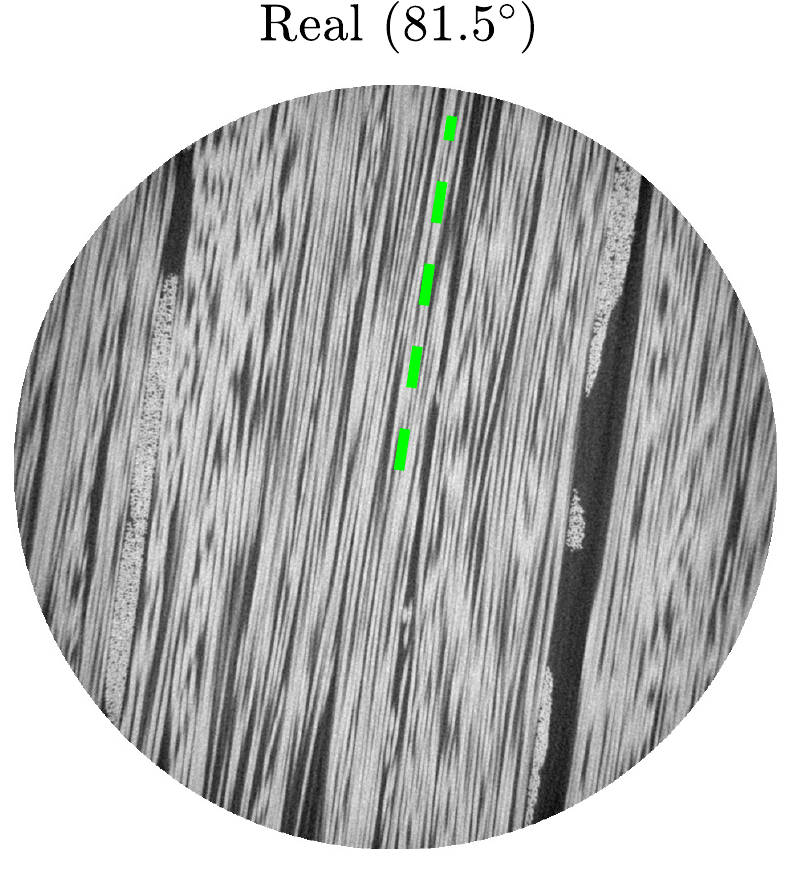}
\hspace{0.05\textwidth}
\includegraphics[width=0.53\textwidth]{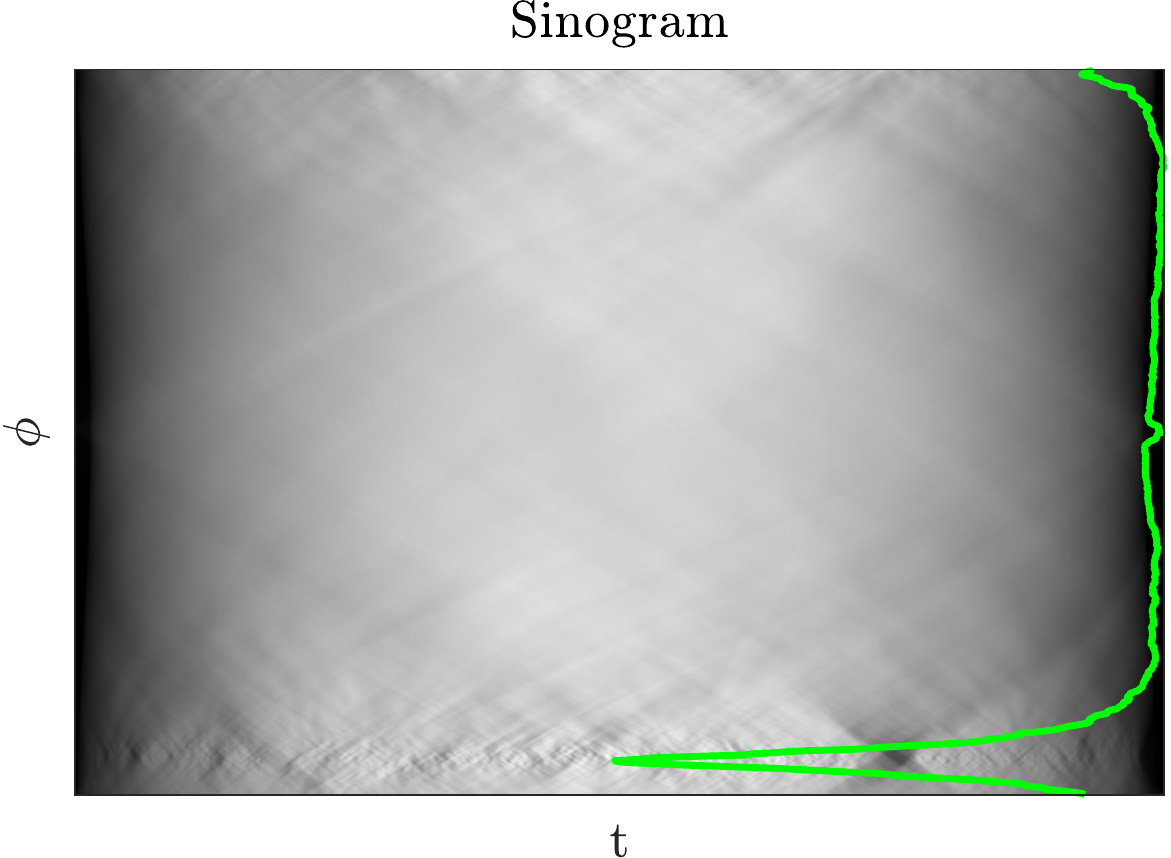}
\caption{To the left: Fibre reconstruction (obtained from \cite{Jespersen2016}, see more in \cite{Jespersen2016b}), with estimated main direction marked by dashed line. To the right: Noise-free sinogram overlayed with a line-plot of $\freqsum(\ctang)$ (see \cref{alg:estang}).} \label{fig:dir_est2}
\end{figure}

We have tested the angle estimation method on a simulated phantom and on a real-data object, where we in both cases simulated the projections and the noise. The real data object is obtained from \cite{Jespersen2016}, for more this data see \cite{Jespersen2016b}. From the noisy sinogram data we have estimated the main direction of the object using \cref{alg:estang}. We tested the method on the two samples with additive Gaussian noise of varying noise-level ($\noisel$) and we report the angle-estimates in \cref{tab:dir_est}. The two tested samples together with their corresponding noise-free sinograms are shown in \cref{fig:dir_est} and \cref{fig:dir_est2}. In these two figures we also indicated the angle estimated from the noise-free sinogram and the angle-dependent summed magnitudes $\freqsum$ from \cref{alg:estang}. 

\begin{table}
\centering
\begin{tabular}{|c|c|c|c|c|c|c|c|c|}
\hline
$\noisel$ (\%) & 0 & 1 & 3 & 5 & 10 & 20 & 30 & 40 \\ \hline
Phantom & 20.1 & 20.1 & 20.1 & 20.1 & 20.1 & 20.1 & 20.1 & 31.7 \\ \hline
Real & 81.5 & 81.7 & 81.5 & 80.9 & 81.7 & 79.5 & -1.1 & -34.9 \\ \hline
\end{tabular}
\caption{Angle estimation results. See phantoms in \cref{fig:dir_est}. For the simulated phantom the exact angle is 20$^\circ$ and for the real data sample we do not have an exact value.} \label{tab:dir_est}
\end{table}

Based on these empirical tests we see that our method is robust up until a noise-level of at least 20\%, which is a high noise-level for CT-data. 


\section{Sinogram Splitting} \label{sec:sinogramsplit}

Edges are essential in the analysis of CT-scans since they indicate material boundaries. An edge is a \textit{singularity}. A singularity is a point where a function, or its derivative, is  discontinuous. From microlocal analysis we know that a singularity is only visible in the sinogram, if an X-ray is tangent to the singularity in the object, see \cite{Quinto1993}. 

In directional objects the texture follows the main direction and therefore the texture edges also follows this direction. The edges along the main direction will be tangent to the X-rays for a specific angle, or within a limited angular range in practical applications. Therefore measurements within this limited range of angles will contain information about the main direction singularities. Based on this observation, we propose a method where the sinogram-data is split into two parts, according to the measurement angles, such that each part will include information about singularities along different directions. For a fibre-crack object, this method could be used to split the data into two parts: one part with singularities of the fibres and one for all other singularities, i.e. cracks etc.

\subsection{Splitting model}

In \cite{Quinto1993}, singularity propagation, from an object through the Radon transform, is examined using \textit{wavefront sets} and microlocal analysis. Wavefront sets denotes points where a given function is not smooth, together with the direction in which the function is not smooth. In \cite{Quinto1993} the relation between singularities in a function $f$ and singularities in the corresponding Radon transform $\radon f$ is described. The paradigm that is described in \cite{Quinto1993} is further outlined in \cite{Krishnan2015} as follows:
\begin{center}
\textit{$\radon$ detects singularities of $f$ perpendicular to the line of integration ("visible" singularities), but singularities of $f$ in other ("invisible") directions do not create singularities of $\radon f$ near the line of integration.}
\end{center}
It should be noted here that a singularity perpendicular to the line of integration corresponds to an edge tangent to the line of integration.

The fact that singularities only propagate when rays are perpendicular to them inspired us to decompose the linear CT-model in separate parts, where each part is related to an object-component where the directions of the singularities, in the object-domain, are limited. The simplest split is a two-component split, which we will base the following on, but the method can be generalized to any integer amount of components with different directions. Components are separated according to the measurement angles: the first part of the model is based on $\splitR+1$ consecutive measurement angles, $\ctangA\in\mathbb{R}^{\splitR+1}$, whereas the second part is based on the remaining angles, $\ctangB\in\mathbb{R}^{\sdim-\splitR-1}$. This splitting results in two linear systems,
\begin{align} \label{eqn:splitmodel}
\sysmA \primfd = \sinoA\qquad \text{and} \qquad \sysmB \secfd = \sinoB,
\end{align}
where $\primfd\in\mathbb{R}^{\pdim^2}$ and $\secfd\in\mathbb{R}^{\pdim^2}$ are the object components. $\sysmA\in\mathbb{R}^{(\splitR+1)\sdim_\ctbin\times\pdim^2}$ and $\sysmB\in\mathbb{R}^{(\sdim-\splitR-1)\times\pdim^2}$ are the split-matrices. And $\sinoA\in\mathbb{R}^{(\splitR+1)\sdim_\ctbin}$ and $\sinoB\in\mathbb{R}^{(\sdim-\splitR-1)}$ are the split measurements. This splitting method is not so suitable for limited data problems, such as limited angle and sparse angle problems, since for such problems we might not have measurements of the singularities that we need in order to be determine $\ctang_\main$. Further more $\splitR>0$ is required to have a reasonable splitting, otherwise the data is simply to limited for the one sub-problem and the reconstruction will be unreliable.

For the example of a split between fibre and crack components, we can split the data and the matrices according to the main object direction $\angE$. 
For main direction $\angE = \ctang_\main$ and even range-width-index $\splitR$, we can pick $\ctangA = (\ctang_{\main-\frac{\splitR}{2}},\ldots,\ctang_\main,\ldots,\ctang_{\main+\frac{\splitR}{2}})$.

This splitting has the downside that objects that do not follow the main direction are very likely to end up in the crack-component, which could complicate a complete fibre-crack segmentation.


Next we introduce two different ways of solving the linear systems in \cref{eqn:splitmodel}. If we use the same linear reconstruction method on both sub-problems we will have have $\objfd=\primfd+\secfd$ and we can therefore analyze both the combined reconstruction and each component individually. On the other hand we can choose a non-linear reconstruction method and use the flexibility of this to impose prior information on each component.

\subsection{Reconstruction methods} \label{sec:sino_reg}

We introduce two different ways to solve the two linear systems \cref{eqn:splitmodel}. The first solution strategy is to use FBP on each system. For the second solution strategy we focus on the variational methods, due to their applicability for inverse problems. This will not necessarily give us a linear reconstruction method, but on the other hand, it will allow us to easily impose some prior information through regularization.

When using standard FBP for limited angle problems, the data from the missing measurement angles is implicitly assumed to be 0, since line integration over 0 will give no impact on the reconstruction. For the two-component split \cref{eqn:splitmodel} this corresponds to the assumptions: $\sysmA \secfd = 0$ and $\sysmB \primfd = 0$. This will result in artifacts which comes from the artificial singularities in the sinogram at the transition between the measured data and the assumed 0-data. For more on limited angle artifacts see \cite{Natterer1986,Frikel2013}. Since FBP is linear, the limited angle artifacts will cancel when summing the two components $\primfd+\secfd$. One can therefore examine the separate components with limited angle artifacts or summed reconstruction without the limited angle artifacts. It should be noted, as also mentioned previously, that the FBP method is not suited for handling underdetermined systems or noise-corrupted data.

Variational methods has been shown to be advantageous for CT-reconstruction problems with limited and/or corrupted data. The motivation for this approach is to solve the sub-problems using prior information about the individual components. So for each component we pick a suitable regularization method and solve the corresponding sub-problem. Formulating the two problems \eqref{eqn:splitmodel} as variational problems and incorporating regularization we get the two optimization problems
\begin{align*}
&\min_{\primfd \geq 0} \| \sysmA \primfd - \sinoA \|_2 + \reguR_\primf(\primfd) \quad \text{and} \\
&\min_{\secfd \geq 0} \| \sysmB \secfd - \sinoB \|_2 + \reguR_\secf(\secfd).
\end{align*}
Both of problems are still limited angle problems, but now the assumptions are only imposed through the regularizers $\reguR_\primf$ and $\reguR_\secf$. Since each component is solved individually, it is possible to incorporate the most suitable regularization technique for each of them.


For the fibre-crack decomposition, the prior for the fibre-component is that it is piecewise constant and that the texture follows one main direction. The prior for the crack-component is that it is independent of the direction, piece-wise constant and sparse. Based on these priors we suggest to regularize the fibre-component using DTV and the crack-component using classical TV together with a $\ell_1$-norm. In discrete terms we have
\begin{align*}
\reguR_\primf(\primfd) = \regup_\primf \DTV(\primfd)  \quad \text{and} \quad \reguR_\secf(\secfd) = \regup_\secf \TV(\secfd) + \sparse \|\secfd\|_1 .
\end{align*}
Both of these regularization choices are convex and due to the piece-wise constant prior they will diminish the undesirable effects of noise.

Solving the sinogram splitting problem using the above suggested regularizers will leave us in total with four tunable parameters: The range-width-index $\splitR$ should be chosen to match the expected range of the fiber direction, i.e. small $\splitR$ for a highly one-directional object and larger $\splitR$ for an object with fibres within a range of angular directions. The regularization parameters $\regup_\primf>0$ and $\regup_\secf>0$ which control the balance between the fidelity terms and the regularizers. And $\secfd$-sparsity parameter $\sparse>0$ which should be relatively small, compared to $\regup_\secf$, in order to avoid $\secfd=0$.

A general advantage of using FBP instead of a variational method is the efficiency, since FBP only requires one single back-projection opposed to several hundred, or thousand, forward- and back-projections when the optimization problem is solved iteratively. General advantages for using a variational method instead of FBP is that artifacts and noise effects can be diminished or removed, while desired features, such as piecewise constancy, can be enhanced.

\section{DTV-decomposition model} \label{sec:dtvdecom}

We wish to impose different priors through regularization on each component in a decomposition model, and hence split the object. This strategy relies on the assumption that the desired components attains a smaller functional value for a desired regularizer opposed to an undesired one.

For an object which consist of some regions that fulfill the directional prior and other regions that \textit{do not} fulfill the prior, we want to develop a model where the two regions are reconstructed separately as different components. Inspired by previous works within texture-cartoon decomposition, see \cite{Aujol2006,Gilles2007}, we opt to build a variational model where the minimizer will be a decomposed reconstruction.


In many applications fibre-structures are analyzed with the aim to detect cracks and/or other types of deterioration. Whereas the texture of the fibre material follows one main direction, the deteriorated parts are mainly perpendicular, or close to perpendicular, to the main direction. Moreover the deteriorated parts are much less dominating in the object, opposed to the fibre materials. Based on these observation we propose the following decomposition model:
\begin{align} \label{eqn:decmod}
\min_{\primf\geq 0,\secf}\ \frac{1}{2}\|\sysm(\primfd+\secfd)-\sino\|_2^2 + \regup \Big( \DTV_{\angE,\widE_\primf}(\primfd) + \reguw\DTV_{\angE^\bot,\widE_\secf}(\secfd) \Big) + \sparse\| \secfd\|_{\ell^1}.
\end{align}
In this model $\primfd\in\mathbb{R}^{\pdim^2}$ represents the fibres and $\secfd\in\mathbb{R}^{\pdim^2}$ the crack part. The summation of these two components is assumed to model the object, hence the summation in the first term of \cref{eqn:decmod}. $\primfd$ is assumed to follow the main object-direction $\angE$, to be piece-wise constant and to be non-negative since it corresponds to attenuation coefficient values. $\secfd$ is assumed to follow the orthogonal main direction $\angE^\bot$, to be piece-wise constant and to be sparse.


The parameters of the decomposition model \cref{eqn:decmod} are the following: The regularization parameter $\regup>0$, which controls the trade off between the regularizer and the fidelity term. The main direction angle $\angE$ which can be estimated with a fast and noise-robust algorithm, see \cref{sec:EstDir}. The first DTV-width $\widE_\primf$ which we assume to be fixed due to previous empirical tests, see \cite{Kongskov2017}. The balance of DTV-terms $\reguw>0$ which should be chosen in combination with the second DTV-width $\widE_\secf\in(0,1]$ to achieve sufficient splitting between the two components. The orthogonal main direction $\angE^\bot$ which is just orthogonal to the main direction. And the crack-sparsity parameter $\sparse>0$, which should be chosen as a relatively small value, compared to $\regup$, to avoid forcing $\secfd$ to be zero. This leaves four tunable parameter: $\regup$, $\reguw$, $\widE_\secf$ and $\sparse$. In order to avoid an overlap of the feasible sets of the two DTV terms we have the following restriction for $\reguw$:
\begin{align*}
\widE_\primf < \reguw < \frac{1}{\widE_\secf}.
\end{align*}
This bound is obtained by comparing the two DTV functionals. We want to avoid both
\begin{align} \label{eqn:normineq}
|\rot_\angE\scm_{\widE_\primf} \exvec|_2 < \reguw|\rot_\angE\scm_{\widE_\secf} \exvec|_2 \quad \text{and} \quad 
\reguw|\rot_\angE\scm_{\widE_\secf} \exvec|_2 < |\rot_\angE\scm_{\widE_\primf} \exvec|_2 ,
\end{align}
for all $\exvec \in\mathbb{R}^2$. Using the matrix definitions in \cref{eqn:dtvmat} and the fact that $\angE^\bot=\angE+\frac{\pi}{2}$ we can rewrite the two norm expression above and achieve:
\begin{align*}
|\rot_\angE\scm_{\widE_\primf} \exvec|_2 &= \sqrt{\hspace{11mm} (\elvec_1\cos\angE + \elvec_2\sin\angE)^2 + \widE_\primf^2(-\elvec_1\sin\angE + \elvec_2\cos\angE)^2 } \\
\reguw|\rot_\angE\scm_{\widE_\secf} \exvec|_2 &= \sqrt{ (\reguw\widE_\secf)^2(\elvec_1\cos\angE + \elvec_2\sin\angE)^2 + \reguw^2(-\elvec_1\sin\angE + \elvec_2\cos\angE)^2 }
\end{align*}
where $\exvec = (\elvec_1,\elvec_2)$. We have $\widE_\primf,\widE_\secf\in ]0,1[$, since we do not want DTV to become classical TV, i.e. $\widE=1$. Based on the two norm-inequalities in \cref{eqn:normineq} we get
\begin{align*}
 1<\reguw\widE_\secf<\reguw \quad &\vee\quad \widE_\primf<\reguw \\
\& \quad \reguw\widE_\secf<1 \quad &\vee\quad  \reguw<\widE_\primf<1 .
\end{align*}
Two of the above inequalities contradict each other and the remaining two gives the bound for $\reguw$
\begin{align*}
\widE_\primf < \reguw \quad\&\quad \reguw\widE_\secf < 1 \iff \widE_\primf < \reguw < \frac{1}{\widE_\secf}.
\end{align*}

The model \cref{eqn:decmod} is convex, which is desirable when we want to find a solution to the minimization problem. Furthermore the sparsity constraint is not only a reasonable regularization method for $\secfd$, it also makes \cref{eqn:decmod} strictly convex, i.e. the minimizer will be unique.


Having introduced two different methods for combined decomposition and reconstruction, sinogram splitting and DTV-decomposition, we sum up the advantages and disadvantages of the two methods: Both methods include four tunable parameters, so in this regard they are similar. 
The sinogram splitting method risks, by design, to reconstruct incorrect attenuation coefficient values. If we choose the DTV reconstruction method for the sinogram splitting, the components are not summable. On the other hand, if we choose the FBP reconstruction method the components are summable, just as for DTV-decomposition where the sum of the components gives the object. The DTV-decomposition method assumes that the crack-component follows a specific direction, where the sinogram splitting method includes no such assumption, only an assumption on the fibre-component.

For the special case where $\sysmA \secfd = 0$ and $\sysmB \primfd = 0$ we have that the data-fidelity terms for the two methods are the same since 
\begin{align*}
\begin{split} \sysmA \secfd = 0 \\ \sysmB \primfd = 0 \end{split} \iff 
\begin{pmatrix} \sysmA \\ \sysmB \end{pmatrix} (\primfd + \secfd) = 
\begin{pmatrix} \sinoA \\ \sinoB  \end{pmatrix} \iff
\sysm (\primfd + \secfd) = \sino \iff
\sysm \objfd = \sino.
\end{align*}

The target for using regularization is also quite different for the two methods. For both methods the regularization is used to suppress noise. The sinogram splitting method furthermore uses the regularization to overcome the limited angle artifact that occur due to the splitting. The DTV-decomposition method utilizes the regularization directly for decomposing the components.

Based on the advantages and disadvantages above, it is not obvious that one method should be superior to the other in all cases, which is why we empirically examine both method in the next section.

\section{Numerical Experiments} \label{sec:numexp}

In this section we demonstrate the performance of the methods introduced in \cref{sec:sinogramsplit} and \ref{sec:dtvdecom}. We do this for a range of simulated X-ray CT problems and we compare the performance of the proposed methods. In order to set the stage for the numerical experiments we first give some discretization and experiment details which are valid for the following tests.

We solve the variational optimization problems using the Primal-Dual-Hybrid-Gradient method from \cite{Chambolle2011a}. 
We stop the algorithm when the relative change of the objective function goes below a threshold tollerance. In all the experiments this tollerance is set to $10^{-5}$. All of the algorithms are implemented in Matlab, where we use the parallel beam GPU code described in \cite{Palenstijn2011} from the ASTRA toolbox, see \cite{VanAarle2015,VanAarle2016}, to calculate forward and backward projections. In all the numerical tests we set the tolerance to $10^{-5}$ and we pick the optimal regularization by ground truth comparison based on the peak signal-to-noise ratio (psnr), unless otherwise stated. We use a simulated crack phantom, which has cracks in a 360$^\circ$ circular pattern, to demonstrate the performance of the previously introduced reconstruction methods. The crack-phantom ground truth can be seen in \cref{fig:crack_phantom}. This phantom does not necessarily fit our assumptions, but serves as a stress-test of the introduced methods.

\figone{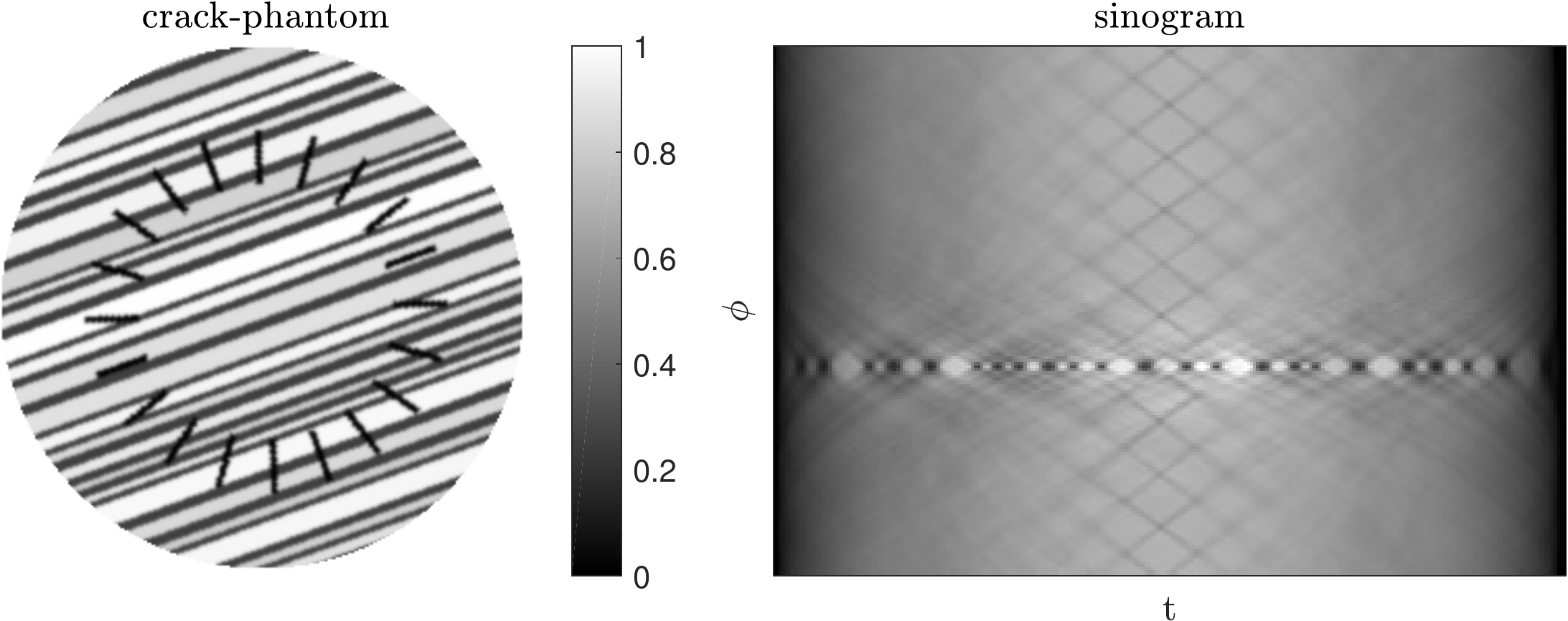}{0.9}{Fibre crack-phantom with fibres along the direction $20^\circ$ and cracks in a circular pattern to the left. Simulated sinogram with no noise shown to the right.}

\subsection{Sinogram splitting}

For the sinogram splitting method we compare the two reconstruction techniques presented in \cref{sec:sino_reg}, namely FBP and the variational method approach. Both reconstruction methods are tested on the same data-set, which is corrupted with 1\% Gaussian noise. The data is simulated with $\sdim_\ctbin=256$ detector bins, $\sdim_\ctang=171$ projections and the reconstruction grid-size is $\pdim=256$. In \cref{fig:split_comp} we compare the two reconstruction methods for a split parameter choice of $\splitR=10$. The choice of $\splitR=10$ is relatively low, which fits well with the phantom being highly directional. 
A comparison between $\primfd$, $\secfd$ or $\primfd+\secfd$ and ground truth $\objfd$ is not suited for picking $\regup_\secf$ since similarity between these signals is not we the model tries to achieve, nor will this yield the most desirable results. When using the variational reconstruction method for sinogram splitting we therefore pick the optimal $\regup_\primf$ and $\regup_\secf$ based on visual inspection instead. The priority for $\primfd$ was to achieve clear edges and an intensity range close to the one in the ground truth. The priority for $\secfd$ was to achieve a reconstruction with a homogeneous background and sharp crack edges.

In \cref{fig:split_comp} we see that both fibre-component ($\primfd$) reconstructions  are visually similar, but the colorbar shows that the intensity level in the FBP reconstruction has an offset of arround 0.5. If we examine the details, the variational reconstruction has much sharper edges and less artifacts than the FBP result. For the crack-component ($\secfd$) reconstructions we see that the noise is very dominating in the FBP reconstruction, whereas the noise is removed by the use of TV-regularization. In general we see that all of the cracks are located in the crack-components and not present in the fibre-components, which is due to a highly directional object and a good choice of the range-width index $\splitR=10$. We even see that the cracks along the main direction are present in crack components. This is due to the edges of the cracks that are perpendicular to the main direction. These edge-singularities will be present in $\sinoB$ and hence the cracks are present in the reconstruction.

\figone{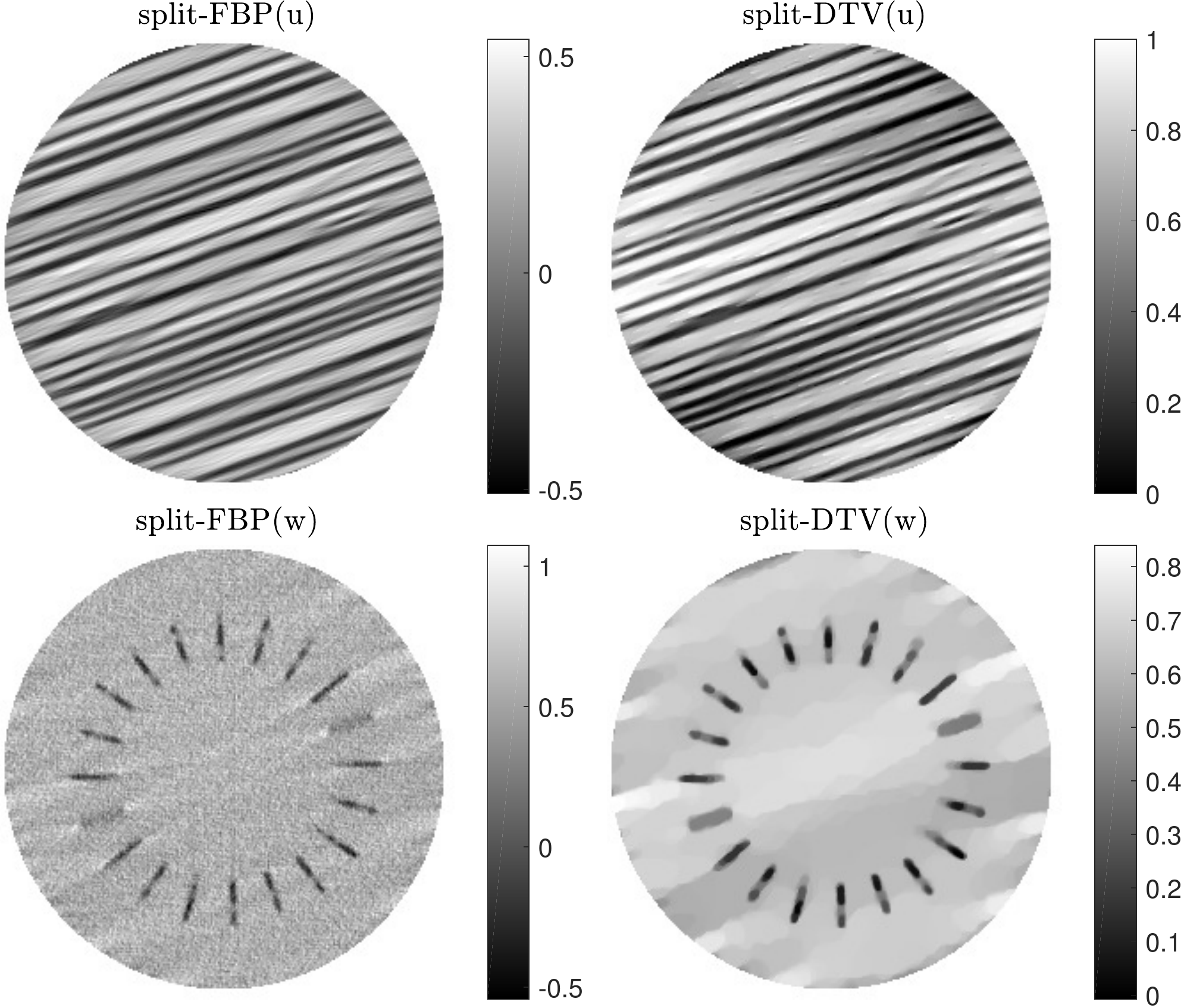}{0.7}{Comparison between two reconstruction methods for the sinogram splitting decomposition technique introduced in \cref{sec:sinogramsplit}.}

To show the role the parameter $\splitR$ we have in \cref{fig:split_K} compared reconstructions using the variational approach for different values of $\splitR$. The test shows that a too low choice of $\splitR$ will result on some fibre elements in the crack-component, whereas too high a choice of $\splitR$ will result in some of the cracks being reconstructed in the fibre-component. The choice of this parameter should be chosen according to prior knowledge about the object, i.e. if the object is highly directional a relatively low value will be sufficient.

\figone{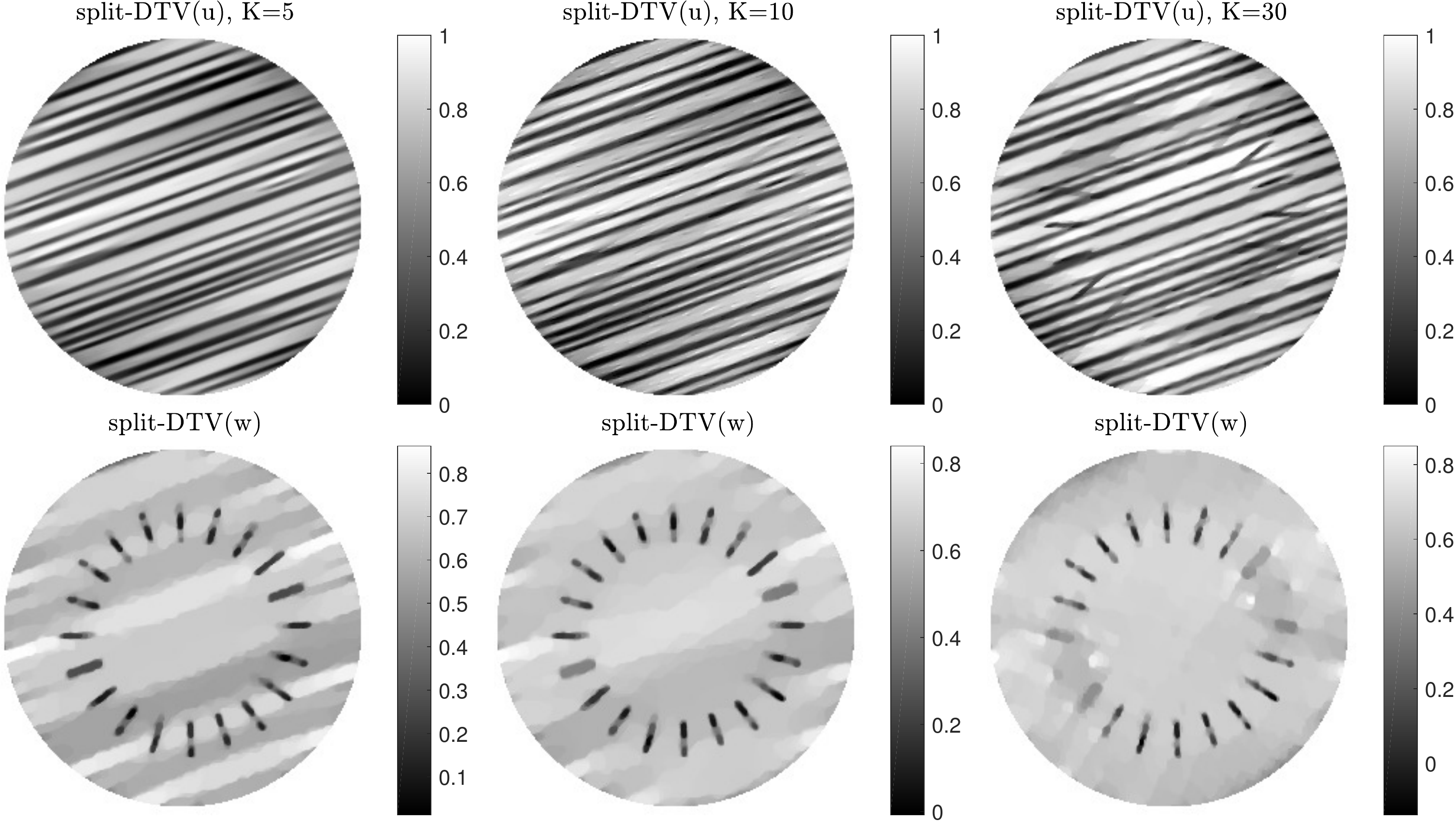}{1}{Parameter-test for the sinogram splitting method using DTV-regularized reconstruction. Reconstructions for different choices of the parameter $\splitR$.}

\subsection{DTV-decomposition}

From the definition of DTV and the empirical tests in \cite{Kongskov2017} we know that the choice the width parameter $\widE$ is related to how well the directional prior fits the given function we try to model. When we use DTV on the crack component $\secfd$ we do not expect all of the crack to follow the direction completely, so we therefore suggest to chose the parameter $\widE_\secf$ higher than $\widE_\primf$, but still lower than 1, to avoid ending up with TV. The value should be chosen according the prior given about the object of interest and for the crack phantom in \cref{fig:crack_phantom} we set $\widE_\secf=0.5$. Based on this choice, together with the already fixed $\widE_\primf=0.15$, we get the bound on: $0.15<\reguw<2$. We pick $\regup$ based on a comparison between the sum of the components $\primfd+\secfd$ and the ground truth $\objfd$. We choose the value of $\regup$ that maximizes the psnr-value.

We tested the influence of the balance between the DTV-terms $\reguw$ and demonstrated some results in \cref{fig:dtv_dec_alpha}. To avoid that the sparsity constraint will influence the results we fix $\sparse=10^{-6}$, which is relatively low, in this test. The visual interpretation of the results in \cref{fig:dtv_dec_alpha} tells us that a low $\reguw$ value will result in more cracks in the the crack component, but also start to introduce noise and fibre-artifact, whereas a high $\reguw$-value will introduce a lot of the cracks in the fibre-component and hence be less desirable, from a decomposition point of view. We observe that the visual optimal choice of $\reguw$ also coincides with the maximum psnr-value based on this observation we used the maximum psnr-value for picking the optimal $\reguw$-value.

\figone{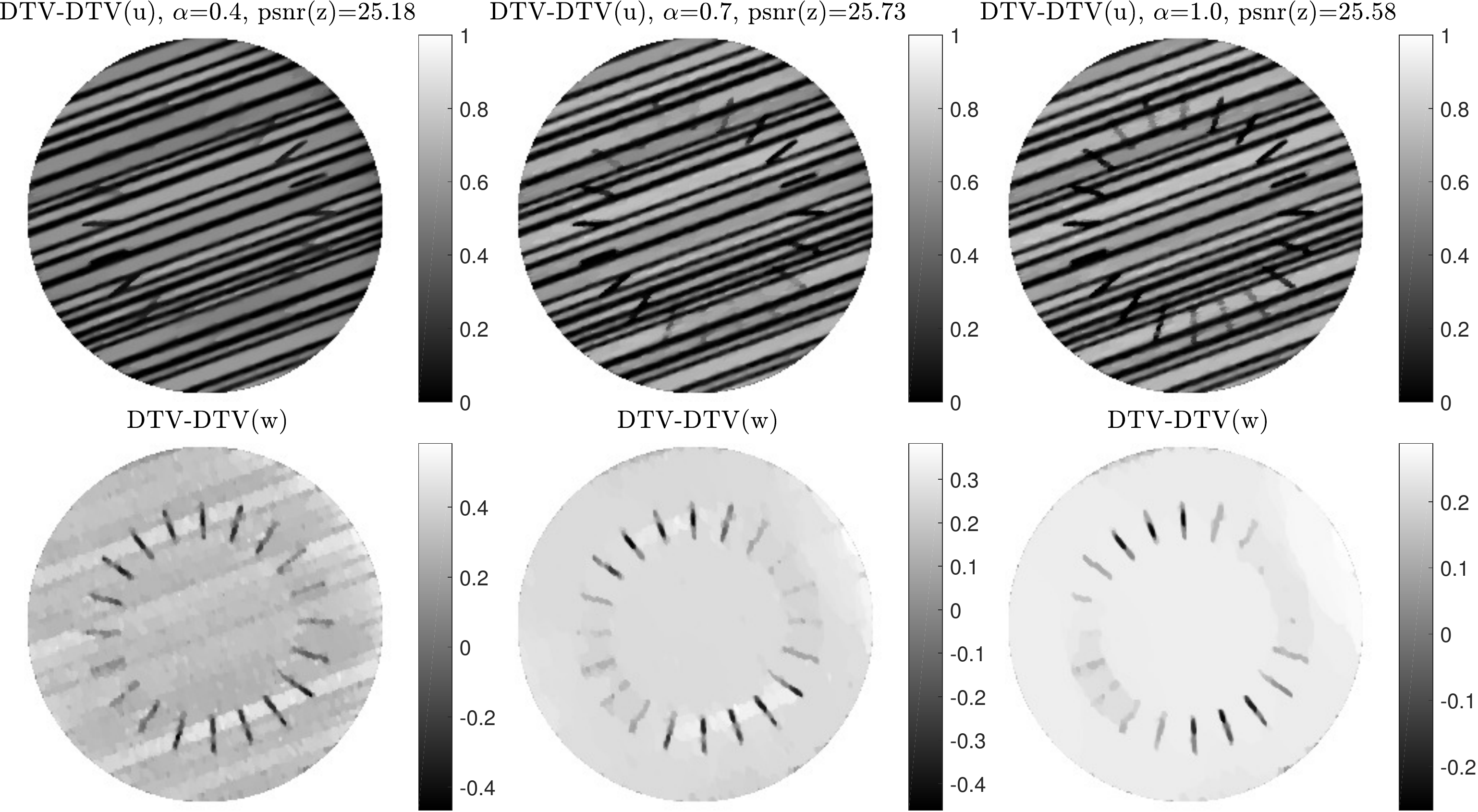}{1}{Parameter test of the DTV balance parameter $\reguw$ for the DTV-decomposition method. Reconstructions for different choices of $\reguw$ visualized.}

We demonstrate the improvement of including the sparsity constraint for the DTV-decomposition method by comparing reconstructions with $\sparse=10^{-6}$ and $\sparse=10^{-4}$. We visualize the the results in \cref{fig:dtv_dec_l1}. From the results we see a clear improvement of both components. The intensity range for the fibre-component is much more accurate and cracks have much sharper edges. The improvement is also reflected by a slight increase of the psnr-value.

\figone{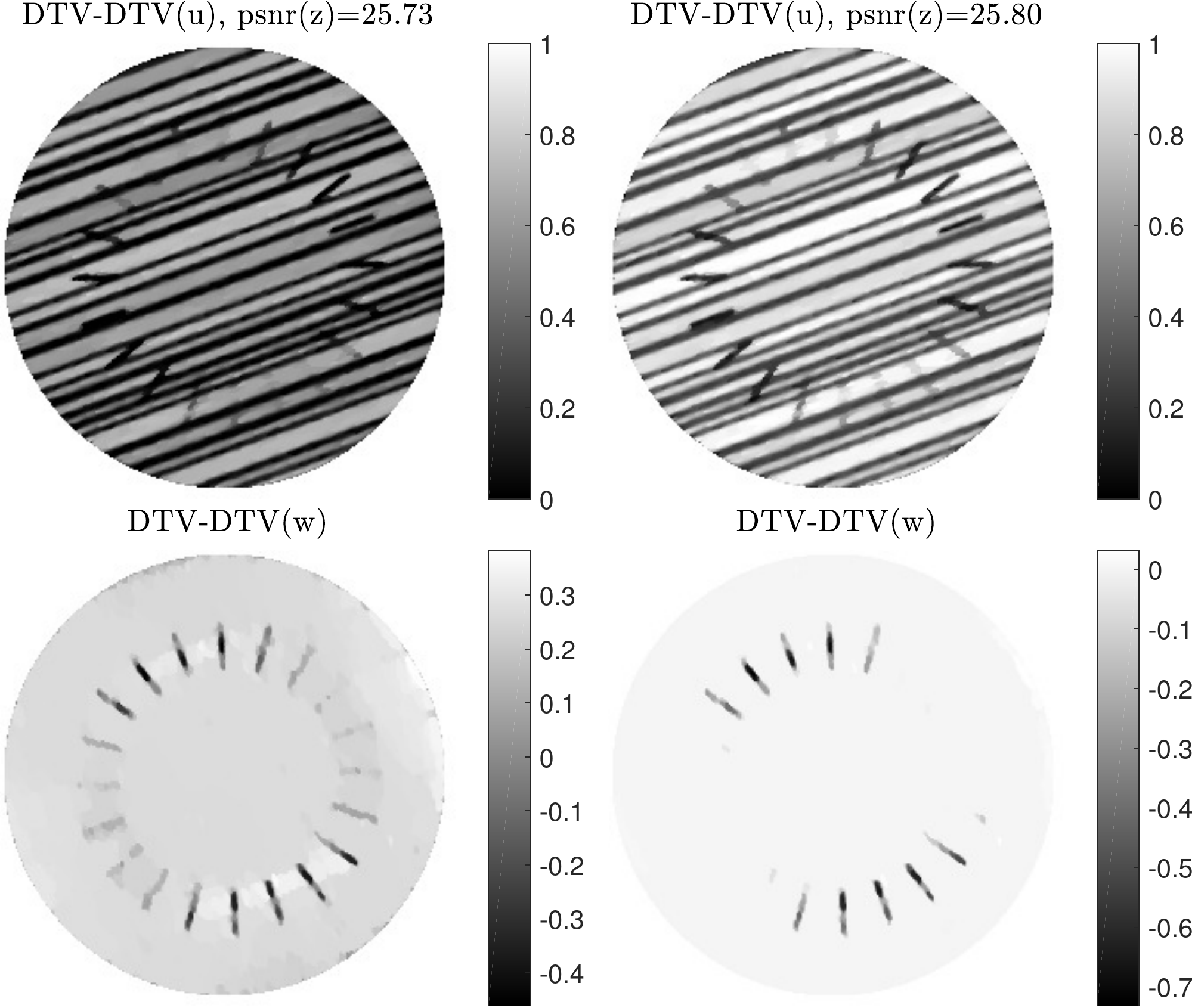}{0.7}{Comparison of the DTV-decomposition method low emphasis on the sparsity constraint, $\sparse=10^{-6}$, and with higher emphasis on the sparsity constraint, $\sparse=10^{-4}$. $\reguw=0.7$ and $\regup=0.0038$. Reconstruction results are visualized.}

\subsection{Sinogram Splitting vs DTV-decomposition}

The demonstration of the two reconstruction methods in the two previous sections shows that the sinogram splitting method delivers a much more complete split between the fibres and the cracks along any given direction. For the sinogram splitting method the crack-component is seen to have a non-homogeneous background, and some 'fibre-artifacts' are present in the regularized reconstruction. Furthermore the cracks appear to be a bit wider than they should be, which is due to the missing data and the strong emphasis which put on the regularizer to remove limited data artifacts and noise. The DTV-decomposition method is best at splitting the fibres from the cracks that are in the range $[\angE+45^\circ,\angE+135^\circ]$ or $[\angE-45^\circ,\angE-135^\circ]$. When the sparsity constraint is enforced the we see that the background of the crack component is highly homogeneous, while the appearing crack-edges are still sharp. Depending of the object and the analysis task on method does not seem superior to the other when testing with this simulated crack-phantom.

Based on the conclusions from the simiulated phantom test we have chosen to compare the sinogram splitting method with the DTV-decomposition method on a real sample object. The carbon fibre sample we have chosen for the comparison can be seen in \cref{fig:real}. The sample is taken from \cite{Rouse2012} where it is analyzed, see page 200-226. We have chosen this sample since it is unidirectional in most of the sample, and that most of the cracks are perpendicular to the main fibre-direction. Based on the sample we simulate data-acquisition as previously and we further simulate noise by adding $1\%$ Gaussian noise to the sinogram. The sample is of size $\pdim=426$ and we simulate $\sdim_\ctbin=426$ detector bins and $\sdim_\ctang=284$ projections.

\figone{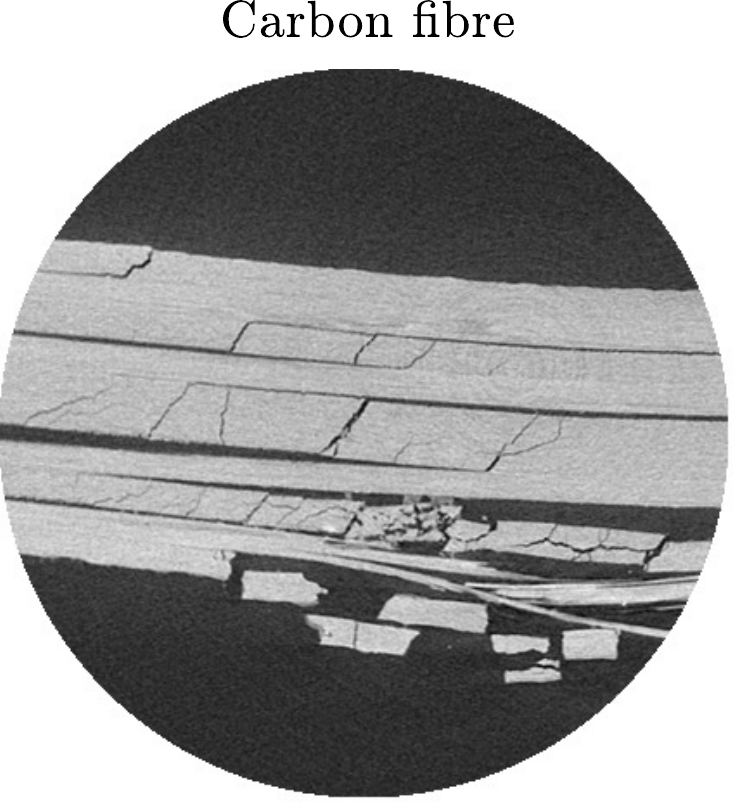}{0.35}{Carbon fibre sample. Reconstruction from CT scan of carbon fibre. Taken from \cite{Rouse2012}.}

In order to have a common reference for the two decomposition methods we have reconstructed the sample using three method that does not decompose the object: FBP, $\ell_2-\TV$ and $\ell_2-\DTV$. The reconstructions can be seen in \cref{fig:comp_real}.

\figone{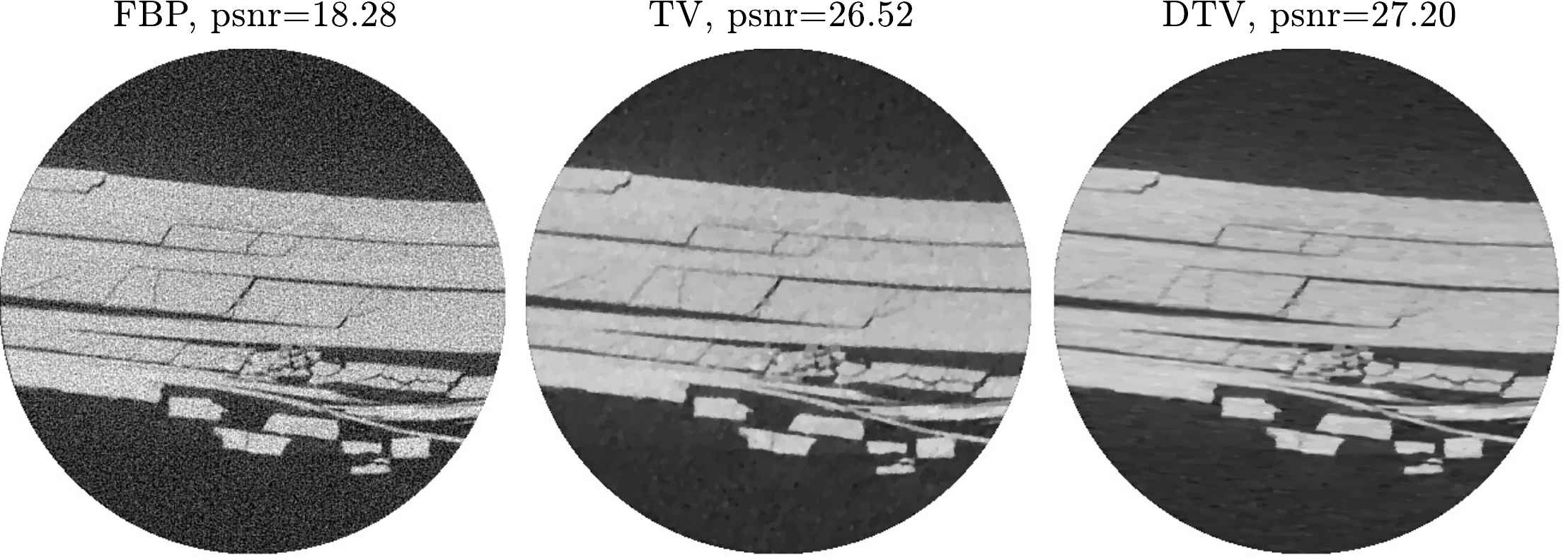}{1}{Simulated CT-problem based on real data object. Solved using three different reconstruction techniques.}

For the sinogram splitting we have tested both FBP and DTV for the reconstruction. We show both the fibre and the crack component for each reconstruction in \cref{fig:comp_real3}. For all three reconstruction methods we tuned the parameters by visual inspection, where we prioritize a decomposition of cracks in one component and non-cracks in the other. Furthermore we tuned the parameters to achieve a homogeneous background for the crack component without smoothing away small detail of the cracks. For the sinogram splitting reconstructions the range-width index $\splitR$ is much larger than for the crack-phantom. This choice is made to avoid having the parts with non-crack singularities, that does not follow the main direction, appearing in the crack-component. By choosing a large $\splitR$ we force the crack-component to have a more homogeneous background.

\figone{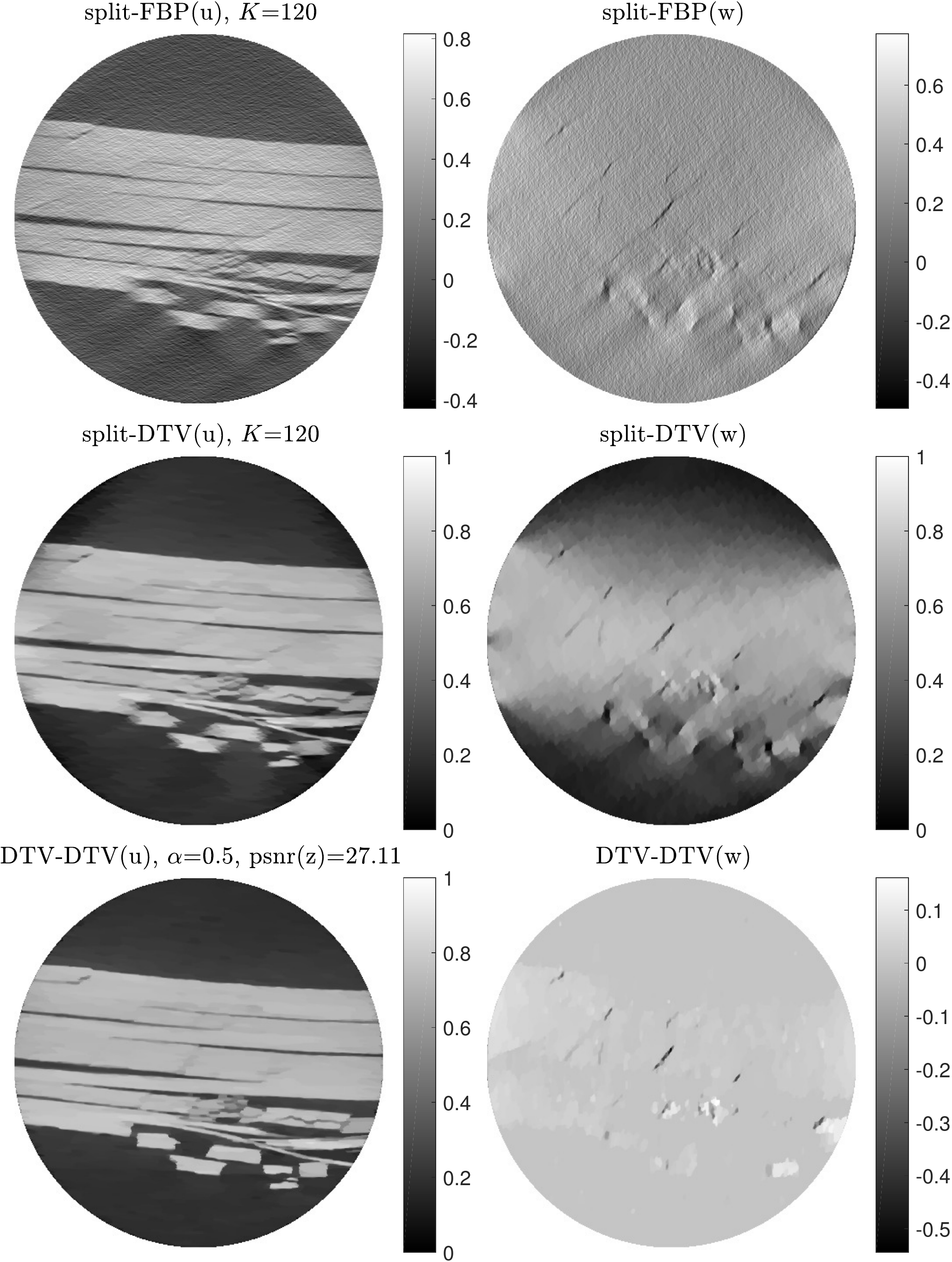}{0.9}{Simulated CT-problem based on real data object. Solved and simultaneously decomposed using three different reconstruction techniques.}

The split-FBP result is seen to be highly influenced by noise in both components and also limited angle artifacts occur. Compared with the two other reconstruction results, the split-FBP method must be seen as inferior. 

The split-DTV result has a sharp edges in the fibre-structure along the main direction, while other edges are blurry, especially the ones perpendicular to the main direction. On the other hand the cracks in the central part of the object are not present in the fibre component as desired. The crack component for the split-DTV method have clearly marked cracks with sharp edges, but this component suffers a lot from stair-casing effects which makes the background highly non-homogeneous. This makes the crack-detection unreliable, since it not certain whether the reconstructed cracks in this component are indeed cracks.

The DTV-DTV reconstruction result has sharp edges for the fibre-component and small-detail parts that could be categorized as cracks are moved to the crack component. The crack-component has a homogeneous background  and sharp crack-edges. Some of the very narrow cracks are not present in the crack-component, but can be observed in the fibre-component.

For this real sample with fibres along one main direction, cracks mostly perpendicular to the fibre-direction and some irregular pieces in bottom right part of sample, the DTV-decomposition has produced the overall best decomposition and reconstruction. The edges are sharper in DTV-decomposition result compared to the split-DTV result, for both components. A disadvantage of the DTV-decomposition is that the narrow crack are not present in the crack-component, but these were already difficult to reconstruct from the underdetermined system with noisy data, cf. \cref{fig:comp_real}.

\section{Conclusions} \label{sec:conclusion}

We have proposed two new tomographic reconstruction methods that makes utilize variational formulations. Both reconstruction methods combine reconstruction and decomposition into one problem and aim to solve both simultaneously. We compare the two method by discussing their theoretical differences. We have also proposed and demonstrated a new method for estimating the main object direction directly from measured computed tomography data. The combined reconstruction and decomposition methods are compared empirically for both a simulated and real data sample. The simulated phantom tests serves as a general performance tests of the methods. In these tests we demonstrate what can be achieved with the proposed methods. The real data sample tests show how well these methods perform in practice. In order achieve even better reconstruction results we consider if the two combined decomposition and reconstruction methods can be combined into a single method.

\section*{Acknowledgements}
The work was supported by Advanced Grant 291405 from the European Research Council.

\bibliographystyle{abbrv}	
\bibliography{DTVDTVproject}

\end{document}